# Numerical Insights into noise amplification of high-energy mid-infrared supercontinuum generation in normal dispersion multimode fibers

Chaofan Yang[†], Dian Duan[†], Fan Zou, Kuo Liu, Ruibo Jin, Zechuan Liu, and Haoyu Wu

*Abstract*—We report on the noise properties of high-energy mid-infrared supercontinuum (MIR-SC) generation in normal dispersion multimode fibers from the numerical perspective. Noise amplification in multi-modes is primarily due to the stimulated Raman scattering (SRS) effect. This leads to the emergence of "incoherent cloud formation" and "incoherent optical wave breaking", similar to those observed in single-mode fibers. Increasing the pump technical noise from 0.1 % to 1 % significantly shortens the lumped coherence length $L_C$ and exacerbates the influence of incoherent broadening dynamics competing with coherent dynamics, resulting in MIR-SC being a strong consistency in the collapse evolution of amplitude noise and phase coherence. To minimize this noise amplification and achieve high-energy low-noise MIR-SC in practical applications, it is essential to use short-pulse pumping with low amplitude noise, ensuring that $L_C \gg L_{OWB}$ (where $L_{OWB}$ denotes the optical wave breaking length).

*Index Terms*—high-energy mid-infrared supercontinuum; noise amplification; multimode fibers; normal dispersion

## I. INTRODUCTION

High-energy mid-infrared supercontinuum (MIR-SC) usually covers the molecular fingerprint region and the atmospheric transparency window, and thus has broad application prospects in the fields of precision laser spectroscopy, atmospheric environment detection, metrology, and national defense and security [1-5]. In recent years there has been a proliferation of research on MIR-SC generation in single-mode optical fibers with fluoride, chalcogenide, and other type of soft glass [6-10], for instance through fiber dispersion engineering, in particular the design of photonic crystal fiber with a special air-hole structure or the use of tapered fibers to achieve zero-dispersion wavelength shift towards shorter wave, thus facilitating efficient and broad-band MIR-SC generation in the anomalous dispersion regime [11-14]. However, since this SC generation dynamics is based on soliton fission, an overload of pulse energy often leads to an excessively large soliton order $N$, such as $N > 10$, causing drastic coherence degradation of SC as well as intensity noise amplification [15], and it usually limits the precision or resolution of systems that use coherence and amplitude fluctuations directly as content in the acquired signal, such as optical frequency comb spectroscopy, optical coherence tomography, or coherent anti-Stokes Raman scattering (CARS) spectroscopy [16-18].

On the contrary, SC generation derived from the coherent spectral broadening dynamics of self-phase modulation (SPM) and optical wave breaking (OWB) in all-normal dispersion (ANDi) single-mode optical fibers can maintain excellent coherent properties, even at high pump powers with narrow pulse widths. In this case the coupling of stimulated Raman scattering (SRS) and four-wave mixing (FWM) leads to a gain saturation, which drastically reduced the gain for the noise-amplifying nonlinear dynamics [19,20]. As a result, even when the soliton order N reaches 500 or more, an octave high-energy SC with low noise characteristics can still be obtained [21]. Therefore, the use of ANDi single-mode optical fibers to generate high-quality MIR-SC has received significant attention.

However, for many mid-infrared applications, higher power spectrum is still especially crucial due to the lack of sensitive detectors. In contrast to a single-mode fiber with a low damage threshold, selecting multimode fibers with larger cores that enable the injection of more power offers more possibilities [22,23]. The typical approach is to utilize a low-loss multi-mode ANDi fiber with step-index or graded-index [24-26] pumped by a multi-kW peak power ultra-short laser. However, another challenging issue arises, that is, at such high powers, the pump laser technical noise will not be negligible, particularly the impact of amplitude fluctuations and pulse duration jitter. Although there are similar considerations for laser technology noise in single-mode ANDi fibers [27], the research on noise amplification of high-energy multi-mode MIR-SC needs further exploration because of the higher peak power in multi-modes and the more complex and multi-dimensional nonlinear coupling between different modes.

In this paper, we numerically investigate the impact of shot noise and technical pump laser fluctuations on the broadening

Manuscript received XX XX, 2024; revised XX, XX; accepted XX, XX. The work was supported in part by the National Natural Science Foundation of China under Grants 11904112, 92365106 and 12074299, in part by Scientific Research Foundation of Wuhan Institute of Technology under Grant K202255, and in part by Natural Science Foundation of Hubei Province under Grant 2022CFA039. *(Corresponding authors: Ruibo Jin; Haoyu Wu.)*

[†]These authors contributed equally to this work and should be considered co-first authors.

Chaofan Yang, Fan Zou, Kuo Liu, Ruibo Jin, Zechuan Liu and Haoyu Wu are with Hubei Key Laboratory of Optical Information and Pattern Recognition, Wuhan Institute of Technology, Wuhan 430205, China (e-mail: witycf97828@163.com; 15727594835@163.com; 17513365784@163.com; jin@wit.edu.cn; lzc647417906@163.com; haoyuwu@wit.edu.cn).

Dian Duan is with Wuhan National Laboratory for Optoelectronics, Huazhong University of Science and Technology, Wuhan 430074, China (e-mail: duandian@hust.edu.cn).

dynamics and noise amplification properties of high-energy MIR-SC generation in multimode fibers at the normal dispersion regime. Although the intermodal energy transfer and nonlinear coupling effects exist in MIR-SC generation, we found that the noise amplification is primarily related to the SRS effect, which leads to the emergence of "incoherent cloud formation" and "incoherent optical wave breaking" like those in single-mode fibers [21]. By increasing the pump amplitude fluctuations from 0.1 % to 1 %, the lumped coherence length in MIR-SC generation will be significantly shortened, contributing to a more drastic SRS effect, thus resulting in a strong correlation collapse evolution of amplitude noise and phase coherence. It is crucial to utilize short-pulse pumping with low amplitude noise to reduce the contribution of incoherent (SRS/FWM) broadening dynamics to facilitate the realization of high-energy low-noise MIR-SC in many practical applications.

## II. SIMULATION MODELS

In order to facilitate the investigation of the impact of pump technical noise, this study is based on the scalar multimode generalized nonlinear Schrödinger equation (MM-GNLSE), which is suitable for multimode fibers with a very large mode area or in cases where the pump laser is singly-polarized, thus circumventing the influence of polarization effects [28-30].

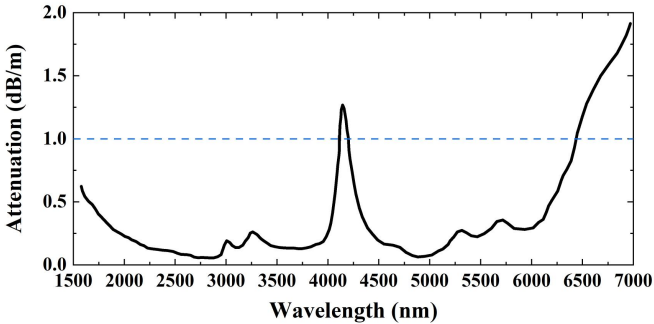

**Fig. 1.** The transmission loss of IRF-S-200 fibers [31].

A step-index commercial multimode As$_2$S$_3$ fiber (IRflex, IRF-S-200) is used here [31]. Its average transmission loss in the spectral range of 1500 nm to 7000 nm is less than 2 dB/m, as shown in Fig. 1. The fractional contribution $f_R$ of the Raman effect to the total nonlinear response is taken as 0.1 with Raman response parameters, $\tau_1$ and $\tau_2$, are 15.2 fs and 230.5 fs, respectively, and the nonlinear coefficient $n_2$ is 1.5×10$^{-18}$ m$^2$/W [32]. Although IRF-S-200 can support hundreds of transverse modes, in MM-GNLSE model, the circular LP$_{0n}$ modes are strongly coupled with the fundamental mode and have a larger overlap with the Gaussian input beam injected to the fiber, thus their contribution to the overall supercontinuum development is much more important than that of other modes. Referring to a recent study [33], only the LP$_{01}$-LP$_{05}$ modes are considered as the primary contributors. By employing finite-element methods, transverse field distributions of the LP$_{01}$-LP$_{05}$ modes and the dispersion curves can be obtained. As shown in Fig. 2, the zero-dispersion wavelength of the LP$_{01}$ mode (the fundamental mode) is about 4.88 μm, while the zero-dispersion wavelengths of the other higher-order linearly polarized modes shift to shorter wavelengths as the order increases.

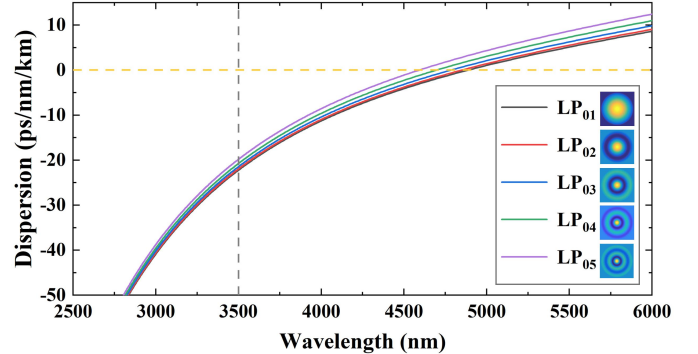

**Fig. 2.** Dispersion curves of LP$_{01}$ - LP$_{05}$ modes (the yellow dashed line corresponds to the zero-dispersion wavelength; the gray dashed line corresponds to the pump wavelength).

The noise model is similar to that of the single-mode ANDi SC [34]. In addition to the typical quantum (shot) noise, the impact of pump technical noise, including amplitude noise and pulse duration jitter are also considered. Therefore, the modified amplitude of the pump laser is expressed as:

$$\tilde{A}_0(\omega) = \mathcal{F}\left\{(1+\psi)\sqrt{P_0}\exp\left[-\left(\frac{t}{(1-0.8\psi)T_0}\right)^2\right]\right\} + \delta_N(\omega), \quad (1)$$

where pump pulse profile is selected as a Gaussian shape, and the first right term carries the technical noise, for a given peak power $P_0$, the peak amplitude of each pump pulse $\sqrt{P_0}$ will be multiplied by $(1+\psi)$, where $\psi$ represents the single random value of each input pulse extracted from a Gaussian distribution with a mean of 0 and a standard deviation equal to the rms amplitude noise of the laser. For the commercial pump laser (OPA laser, OperASolo), the standard deviation is typically taken from 0.1 % to 1 %. And since the pulse duration is inversely proportional to its amplitude, the amplitude noise induces much jittering of the pulse duration, causing the pulse duration to change from $T_0$ to $(1-0.8\psi)T_0$. The second right term models the well-known quantum (shot) noise whose detailed expression can be found in Ref. [34], and it adds a noise seed at the input of one-photon-per-mode (OPPM) with a random phase on each frequency bin. By compiling every one of the 190 possible pairs of spectra (i.e., 20 rounds of simulation), we can compute the modulus of the complex degree of first-order coherence $|g_{12}^p(\omega)|$ for each circular mode. And for the purpose of the following analysis, the lumped coherence $|g_{12}(\omega)|$ with a mean of all circular modes, and the spectrally averaged coherence $\langle|g_{12}|\rangle$ are defined as [34]:

$$|g_{12}(\omega)| = \frac{1}{m}\sum_p^m |g_{12}^p(\omega)|$$

$$= \frac{1}{m}\sum_p^m \left|\frac{\left\langle\tilde{A}_{pi}^*(\omega)\tilde{A}_{pj}(\omega)\right\rangle_{i\neq j}}{\sqrt{\left\langle|\tilde{A}_{pi}(\omega)|^2\right\rangle\left\langle|\tilde{A}_{pj}(\omega)|^2\right\rangle}}\right|, \quad (2)$$

$$\langle|g_{12}|\rangle = \frac{1}{m}\sum_p^m \frac{\int_0^\infty |g_{12}^p(\omega)|\langle|\tilde{A}_p(\omega)|^2\rangle d\omega}{\int_0^\infty \langle|\tilde{A}_p(\omega)|^2\rangle d\omega}. \quad (3)$$

Additionally, we also consider the relative intensity noise $RIN_p$ of the multimode MIR-SC under different pump technical noise. And the lumped relative intensity noise $RIN$ and the spectral range-averaged relative intensity noise $\langle RIN \rangle$ are also defined as [29]:

$$RIN(\omega) = \frac{1}{m}\sum_p^m RIN_p(\omega)$$

$$= \frac{1}{m}\sum_p^m \frac{\sqrt{\langle(|\tilde{A}_p(\omega)|^2-\langle|\tilde{A}_p(\omega)|^2\rangle)^2\rangle}}{\langle|\tilde{A}_p(\omega)|^2\rangle}, \quad (4)$$

$$\langle RIN \rangle = \frac{1}{m}\sum_p^m \frac{\int_0^{edge} |RIN_p(\omega)|\langle|\tilde{A}_p(\omega)|^2\rangle d\omega}{\int_0^{edge} \langle|\tilde{A}_p(\omega)|^2\rangle d\omega}. \quad (5)$$

## III. RESULTS AND DISCUSSION

### A. The feasibility of simulation models

To verify the validity of the scalar MM-GNLSE model, we examined the simulated SC generation using the same fiber and pump laser parameters as the experiments in the literature [31]. This involved assuming a Gaussian-shape pump pulse with a center wavelength of 3500 nm, a pulse duration of 150 fs, and a peak power of 3 MW, and it is injected into an IRF-S-200 fiber of 1 m length for MIR-SC generation, considering only quantum noise. At the end of IRF-S-200 fibers, the simulated SC is in general agreement with the experimental results of reference [31], as shown in Fig. 3(a). The modulation structures at 3580 nm and 3757 nm near the pump wavelengths have a close correspondence, and the spikes observed at the spectral edges at 4310 nm and 4715 nm also match well. In addition, at the short-wave 2500 nm band, the simulated spectral results reproduce a dense modulation structure consistent with the experimental results.

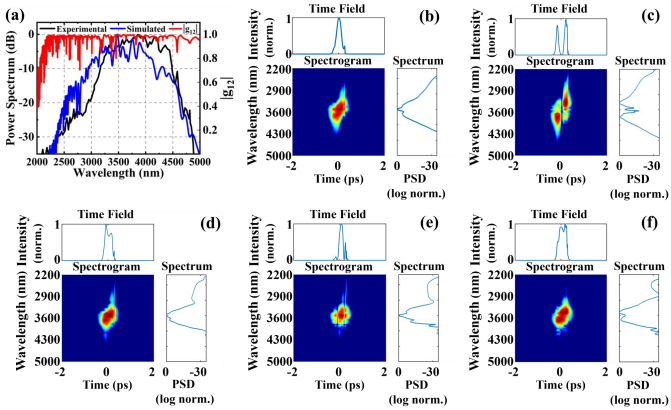

**Fig. 3.** (a) Comparison of numerical simulation SC with experimental results in reference [31]; (b)-(f) when the propagation distance is 0.016 m, the temporal profiles and normalized logarithmic spectrum and spectrograms of the $LP_{01}$-$LP_{05}$ modes in sequence.

From the simulation for different modes, we find that the evolution of multi-mode SC is more complicated. Since the nonlinear length of the modes $L_{NL} = \lambda/(3\pi n_2 S_{plmn}^K P_0)$ is much smaller than the intermodal walk-off length $L_W^{pq} = T_0/|\beta_1^{(p)} - \beta_1^{(q)}|$, where $S_{plmn}^K$ is the overlap integral of the individual modes, there is energy transfer and intermodal cross-phase modulation (XPM) between the $LP_{0n}$ modes [28]. As shown in Fig. 4, within the propagation distance of 0.05 m, it is evident that modes of $LP_{01}$ and $LP_{02}$ have undergone strong energy transfer, leading to a substantial impact on the evolution of pulse shapes. That is, the pulse temporal profiles for two modes exhibit alternating periodic breaking and reshaping, specifically can be seen at the length from 0.008 m to 0.022 m. At distances of 0.016 m and 0.018 m, we also observe an interference in the tail of the temporal profile of the pulse for both modes. This is because the fiber dispersion at shorter wavelengths is steeper than that at longer wavelengths. Coherent OWB that generates new short wavelength components has occurred at the trailing edge of the pulse, but has not yet begun at the pulse front, as shown in Fig. 3(b-c).

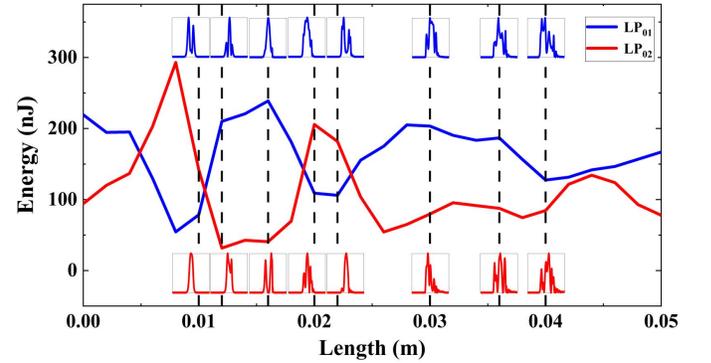

**Fig. 4.** Energy transfer between $LP_{01}$ and $LP_{02}$ modes at a propagation distance of 0.05 m (solid line) and evolution of pulse shape of two modes at different propagation distances (insets).

As the pulse laser continues to propagate, we observe a rapid temporal broadening with decreasing peak power caused by the fiber dispersion, resulting in a reduction of nonlinear effects between the $LP_{01}$ and $LP_{02}$ modes. When the distance exceeds the intermodal walk-off length (i.e. 0.302 m), the spectra of the two modes will evolve almost independently. Similar evolutions also occur in the SC generation of higher-order linearly polarized modes. But here SC generates in all-normal dispersion regime and will still be affected by SPM, OWB and SRS. So returning to the Fig. 3(a), the combination of SPM and OWB results in the total spectral modulation structures near the pump wavelengths at 3580 nm and 3757 nm, as well as the spikes at the edge of the long wave. Additionally, the nonlinear coupling of SRS to the parametric FWM produces anti-Stokes peaks, which likely exacerbates the spectral modulation structure at 2500 nm, and the anti-Stokes peaks will further lead to the deterioration of SC phase coherence, which can also be observed in Fig. 3(a) that the coherence $|g_{12}(\omega)|$ in the band near 2500 nm exhibits many glitches (red line).

### B. The combinatorial impact of technical pump laser fluctuations and SRS effect on SC noise amplification

In the following, we introduce the pump technical noise to investigate variations of the average spectral coherence $\langle|g_{12}|\rangle$.

By fixing the pulse peak power $P_0 = 2$ MW with amplitude noise $\psi = 0.3$ %, the average spectral coherence as a function of pump pulse duration and propagation distance is plotted in Fig. 5(a). Lighter-colored areas indicate higher spectral coherence, while darker-colored areas signify the opposite. When the input pulse duration is less than 300 fs, it can achieve an average coherence of over 0.9 (gray solid line) across the entire 1 m fiber length. However, if the pulse width is larger than 300 fs, at the same propagation distance, the SC coherence will degradation dramatically as the pulse duration continues to increase. When the pulse duration is increased to 1 ps, the SC will quickly decohere after the propagation distance exceeds 0.1 m. Thus, for long pulse durations, shortening the propagation distance can ensure sufficiently high spectral coherence, but at the expense of spectral width.

Fig. 5(b) illustrates the $\langle |g_{12}| \rangle = 0.9$ contour results for the case of a pump pulse with 0.1 % -1 % amplitude noise, and the contour lines move toward shorter pulse widths as the pump amplitude noise increases. Under the condition of maximum 1 % amplitude noise, the high coherence area (i.e. light-colored areas as shown in Fig. 5(a)) for a broadband MIR-SC will require pulse widths below 200 fs. This shows that pump technical noise exerts a significant limiting effect on the phase coherence.

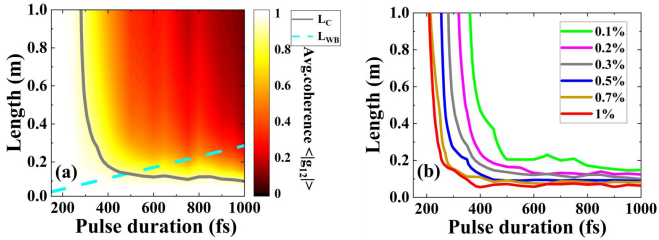

**Fig. 5.** (a) Average spectral coherence of SC generated with 2 MW peak power pump pulses as a function of pump pulse duration and propagation distance including a technical amplitude noise value of 0.3 %. The gray solid line indicates where the spectrally averaged coherence is 0.9. The cyan dashed line indicates the lumped OWB length. (b) Evolution of the spectrally averaged coherence value of 0.9 for various levels of technical amplitude noise going from 0.1 %-1 %.

In order to elucidate the evolutionary dynamics of the above coherence properties, we have employed feature lengths comparable to those in the single-mode ANDi-SC for analysis [21]. Due to the selection of five transverse modes $LP_{01}$-$LP_{05}$, we have calculated the average feature length of all modes lumped together. The spectrogram for different modes are also superimposed and averaged in the following analysis. Note that although this treatment is not rigorous due to nonlinear mode coupling, we have found that it leads to very meaningful conclusions.

Still considering the above case of $\psi = 0.3$ %, we define a lumped soliton order $N = \sum_p^m \sqrt{L_D^p / L_{NL}^p}$, where $L_D^p$ and $L_{NL}^p$ represent dispersive length and nonlinear length respectively. For pulse durations of 0.05 ps to 1 ps, the lumped soliton order $N$ ranges from 7.37 to 147.50. Herein, the feature lengths are also offered, that is, the lumped OWB length is given by $L_{WB} \approx (1.1/m) \sum_p^m \sqrt{L_{NL}^p L_D^p}$ and the lumped coherence length $L_C$ is defined as a propagation length for the SC spectrally average coherence $\langle |g_{12}| \rangle = 0.9$, corresponding to the cyan dashed line and the gray solid line in Fig. 5(a) respectively.

For understanding evolution of spectral coherence of the high-energy MIR-SC generation, we will perform the elaborate analysis in the cases when $L_{WB} < L_C$ and $L_{WB} > L_C$; such as when the pulse duration $T_0 < 350$ fs, where $L_C \gg L_{WB}$, it can be clearly observed in Fig. 5(a) that the phase coherence of SC is high enough, although the energy transfer between different modes and intermodal XPM occurs, the different modes of SC generation are mainly based on coherent (SPM/WB) spectrum broadening.

When the pulse duration satisfies 350 fs $< T_0 <$ 480 fs, $L_C$ decreases rapidly and becomes comparable to $L_{WB}$. The lumped SC spectra and the coherence $|g_{12}(\omega)|$ evolution of a 2 MW, 450 fs pump laser pulse including an amplitude fluctuation of 0.3 % are summarized in Fig. 6. It can be observed that at propagation distances $z$ less than 0.06 m before any significant coherence degradation, the short wavelength of coherent SC is fully formed, while the long wavelength is suppressed due to the stronger transmission loss in the $As_2S_3$ fiber. After $z > 0.1$ m [position I], the noise seed SRS feature emerges at the temporal peak of the pulse as shown in Fig. 6(c). Moreover, as the much steeper fiber dispersion curve for short-wavelength, OWB occurs at the trailing edge of the pulse rather than the leading edge.

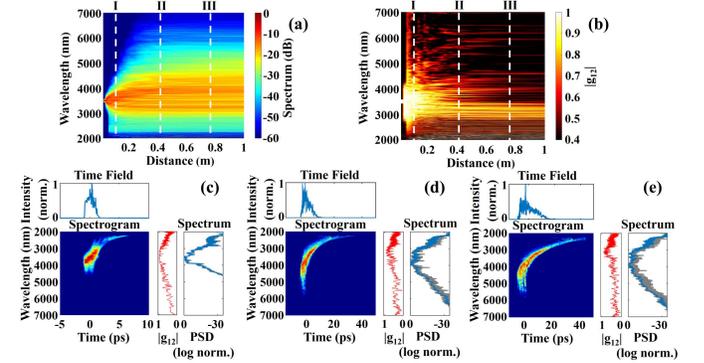

**Fig. 6.** Evolution of (a) lumped spectral intensity and (b) spectral coherence of a 2 MW, 450 fs pump laser pulse including a technical amplitude noise value of 0.3 % over a propagation distance of 1 m, (c-e) lumped spectrogram where projected axes show spectral coherence (red), mean spectrum (blue), spectral fluctuations (gray) and the temporal pulse profile at propagation distances marked I, II, and III in (a) and (b) corresponding to 0.1 m, 0.42 m, and 0.78 m, respectively. $z$ indicates the direction of propagation.

As the propagation continues [position II], the SRS component spread towards the peak and leading edge of the pulse, thereby facilitating the expansion of the long-wavelength spectrum in Fig. 6(d). Simultaneously, the Raman assisted parametric FWM rapidly propagates the noise into the shortwave band. The presence of the SRS and FWM components, along with their overlap with the main pulse cause noisy interference structures and large shot-to-shot fluctuations in both the frequency and temporal domains.

These dynamics can lead to a progressive coherence degradation, starting from the outermost blueshifted SPM peaks spreading towards the spectral edges, predominantly towards longer wavelengths, observed in the spectral fluctuations of the 20 rounds of simulation in Fig. 6(d) (i.e. gray spectral fluctuations). As the pulse continues to propagate more [position III], the SC spectral shape does not change significantly as seen in Fig.6(e), but coherence degradation continues to occur in some components of the spectrum, directly leading to a broader dip around the 5000 nm wavelength band. And the corresponding single-shot lumped spectrogram exhibits an incoherent and noisy "cloud" forms.

Then when the pulse duration $T_0 > 480$ fs and $L_C$ is shorter than $L_{WB}$, and a 2 MW, 650 fs pump pulse including the fixed amplitude noise value $\psi$ of 0.3 % is taken, the lumped SC and coherence $|g_{12}(\omega)|$ evolution are shown in Fig. 7. It is evident that the phase coherence feature evolution closely resembles of Fig. 6, especially before the distance of 0.4 m. The same noise seed SRS feature emerges as $z > 0.06$ m [position I], corresponding to the highest wavelength intensity in the temporal domain, and its component spreads towards the peak and leading edge of the pulse, resulting in the broadening of the long-wave. Simultaneously the SRS/FWM decoherence mechanism already discussed in Fig. 6 is active as well, leading to noisy interference structures in both the frequency and time domains [position II].

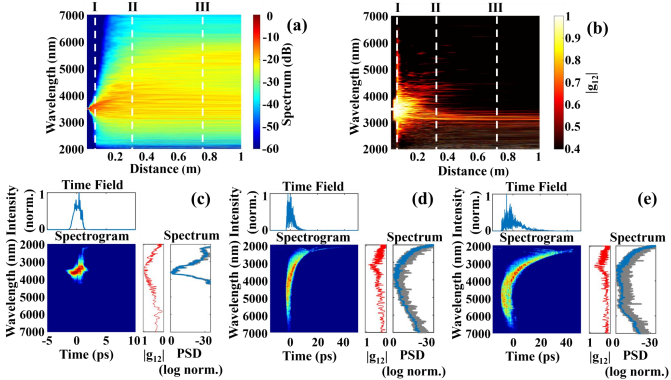

**Fig. 7.** Evolution of (a) lumped spectral intensity and (b) spectral coherence of a 2 MW, 650 fs pump pulse including a technical amplitude noise value of 0.3 % over a propagation distance of 1 m, (c-e) lumped spectrogram where projected axes showing spectral coherence, mean spectrum (blue), spectral fluctuations (gray) and the temporal pulse profile at propagation distances marked I, II, and III in (a) and (b) corresponding to 0.06 m, 0.32 m, and 0.76 m, respectively.

However, as the pulse continues to propagate [position III], since the dispersion of the long-wavelength band is much larger than short waves, the Raman redshifted long-wavelength pulse overlaps with the SPM component and the main pulse front, resulting in the process known as "incoherent optical wave breaking" [21]. This is similar to coherent OWB for allowing the transfer energy from the central part to the spectral wings, which results in a dip in the center spectrum as shown in Fig.7(e). Whereas, due to the impact of gradual noise-seeded SRS frequency generation, the long wavelength edge is incoherent. And the short wavelength at around 3000 nm is contained in a long temporal tail and propagates at the slowest velocity in the back of the pulse, it exhibits low temporal intensity and is limitedly affected by SRS/FWM. Consequently, the phase coherence at around 3000 nm keeps relatively high.

As a result, the analysis of above two cases revealed that the SRS effect significantly influences the noise amplification of the multi-mode SC, leading to the emergence of "incoherent cloud formation" and "incoherent optical waves breaking". Thus, it is very necessary to reduce the role of incoherent (SRS/FWM) broadening dynamics in competition with coherent (SPM/WB) dynamics for MIR-SC generation in multimode ANDi fibers. An effective approach is to use ultra-short pump pulses. However, when the pump amplitude noise is increased only from 0.3 % to 0.5 %, the lumped coherence length $L_C$ of SC generation will be significantly shortened, resulting in such a drastic SRS effect that the use of a pulse duration of 300 fs is no longer effective. Therefore, in this case, to achieve a broadband and highly coherent SC, the pump laser pulse duration should be shorter to ensure that $L_C \gg L_{WB}$.

Finally, we also investigate the impact of the pump technical noise on the *RIN* of multimode MIR-SC. To reduce SC noise coupling between phase and amplitude, a pulse duration of 100 fs was selected to ensure a sufficiently high average coherence $\langle |g_{12}| \rangle$ of that is greater than 0.9. For several pump amplitude noise of 0.1 %, 0.3 %, and 0.5 %, the *RIN* of MIR-SC after 1 m of propagation are shown in Fig. 8(a). It illustrates that the center spectrum of MIR-SC exhibits significantly lower *RIN* compared to the both edges, and even show a remarkable resistance against pump technical noise, which is also demonstrated in recent multimode MIR-SC experiments [35]. As the technical noise increases, the *RIN* profiles bandwidth of the noise below the pump will be sharply narrower as the *RIN* at both edges continues to rise. Notably, the *RIN* in the short-wavelength edge is much higher than the long ones, we attribute it to the higher transmission loss in the band near 4100 nm and beyond 6000 nm in $As_2S_3$ fibers, which provides a stronger filtering to suppress the *RIN* in long-waves.

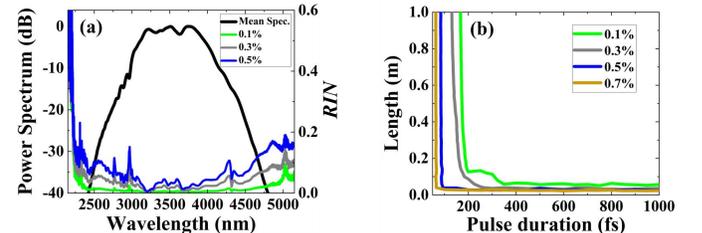

**Fig. 8.** (a) *RIN* profiles for various levels of pump technical noise (color curves) and the mean spectrum obtained from 20 rounds of simulation after 1 m of IRF-S-200 fiber with a 2 MW, 100 fs pump pulse including a technical amplitude noise value of 0.5 %. (b) $\langle RIN \rangle = 2$ % contour plots for various levels of pump technical noise going from 0.1 %-0.7 %.

The properties of the averaged spectrum intensity noise $\langle RIN \rangle$ is shown in Fig. 8(b). $\langle RIN \rangle = 2$ % contours shift towards shorter pulse durations as the pump amplitude noise increases, which is similar to the average coherence $\langle |g_{12}| \rangle$ contours in Fig. 5(b). The reason for this important finding can be attributed to the reduction in lumped coherence length, thus contributing to a more drastic SRS effect, leading to a strong collapse evolution of amplitude noise consistent with phase coherence. Carefully choosing a shorter propagation distance can reduce the impact of SRS to ensure a low $\langle RIN \rangle$. But if pump amplitude noise $\psi$ gradually approaches 1 %, the contour here moves out of the pulse width range, making it impossible to achieve $\langle RIN \rangle = 2\ \%$ for even shorter propagation distances.

## IV. Conclusion

In summary, we have shown quantitatively the noise amplification properties of high-energy MIR-SC in normal dispersion multimode fibers with several input pump noise. The increased pump technical noise will shorten the lumped coherence length and exacerbate the explicit influence of SRS, thereby making MIR-SC a strong consistency in the evolution of amplitude noise deterioration and coherence degradation. But during this collapse evolution, a few components of the SC can still maintain sufficiently high coherence, or their intensity noise is even lower than that of the pump laser. This remarkable resistance property is very interesting. In addition, the simulation results also demonstrate that it is critical to reduce the role of incoherent (SRS/FWM) broadening dynamics in competition with coherent (SPM/WB) dynamics by using a short-pulse pump with low amplitude noise, and in this case, to obtain the high-quality SC, it is usually required to ensure that $L_C \gg L_{WB}$. Our research findings may be also applicable to the graded-index multimode fiber to ensure that the generated MIR-SC not only has a large energy and low-noise, but also benefits from stronger mode cleaning effect caused by the graded refractive index, thereby further enhancing the spatial coherence.